# Perceptions of Task Interdependence in Software Development: An Industrial Case Study

Mayara Benício de Barros Souza
*Universidade Federal de Pernambuco and Universidade Federal do Vale do São Francisco*
Brazil
mbbs@cin.ufpe.br

Fabio Q. B. da Silva
*Universidade Federal de Pernambuco*
Brazil
fabio@cin.ufpe.br

Carolyn Seaman
*UMBC*
USA
cseaman@umbc.edu

*Abstract*— **Context**: Task interdependence is a work design factor that expresses the mutual dependency between tasks that compose a whole work. In software development, task interdependencies are created by the technical dependencies between the components of the software system and by how the development tasks are allocated to individuals in a teamwork context. Despite its importance for individual and team effectiveness, we still do not have studies about how software engineers perceive task interdependence in practice. **Goal**: To understand the perceptions of software engineers about the interdependence in their work and how these perceptions interact with other human and technical factors in the development process. **Method**: We performed an exploratory qualitative case study of a single software development team in a Brazilian software company that developed solutions for the financial market. We interviewed all 10 team members and used standard coding techniques from qualitative research to code, categorize, and synthesize data. **Results**: Individuals are consistent in their understanding of task interdependence and how it happens in practice. However, there are asymmetries between the individual perceptions in an interdependence relationship, which seem to exacerbate expressed feelings of anxiety and dissatisfaction. **Conclusion**: Our results suggest that the perception of task interdependence in software development is often not symmetrical with potential negative effects on emotional states that are related to motivation and satisfaction in the workplace.

*Keywords— task interdependence, work design, qualitative research, case study, software engineering*

## I. Introduction

Task interdependence expresses the mutual dependency between tasks that compose a whole work i.e., the degree to which a task depends on others and others depend on it for the success of the work [1][2]. Interdependence expresses the ways different tasks that compose a whole work or project are interrelated. Research in work design has found that interdependence correlates with other work-related factors. For instance, Kiggundu [1] and Grabner et al. [3] contended that task interdependence positively correlates with job satisfaction. The latter also studied the trade-offs between individual autonomy and task interdependence in the context of creative teams, which share similarities with software teams.

Despite the potential motivational benefits defended by Kiggundu [1] and supported by Morgeson and Humprhey [2], Silva et al. [9] found evidence that interdependent tasks correlate with an increase in perception of role conflict in software teamwork. Also, Marsicano et al. [10] identified that in software development task interdependence increases the need for information sharing and synchronization of tasks. Moreover, according to Janz et al. [11] and Kakar [4], high levels of interdependency between software team members may be perceived as counterproductive because the time spent on negotiations and decision making processes could be used to actually carry out the tasks. Moreover, Kakar [4] found that high levels of task interdependence may induce an increase in team cohesion, but are likely to decrease individual autonomy for performing tasks, which is an important work design motivational factor [2][12], in particular in agile software development [13].

Software development is well known to require the interplay of different technical and managerial roles, and the use of a diversity of skills, mostly in a teamwork situation [4][5][6][7]. Thus, teamwork design in software development entails high levels of task interdependence, created by the technical dependencies between the components of the software system (system interdependence) and also by how the development tasks are allocated to individuals in a teamwork context (process interdependence). Based on Conway's Law, we could also contend that the way the software organization is structured also affects the interdependence of software development tasks (organizational interdependencies) [8].

System interdependencies are part of the software project complexity and, as such, cannot be completely eliminated although good design techniques may reduce them. On the other hand, process and organizational interdependencies could be handled if properly identified and explicitly managed.

Thus, task interdependence in general is an unavoidable, but possibly manageable job design characteristic in software development that may have positive or adverse relationships with other technical and human related factors in the work. Due to its importance for individual and team effectiveness, in depth studies of the perception of software engineers about task interdependence are needed. Therefore, the goal of this study is to investigate the perceptions of software engineers about task interdependence and its relationships with other human and technical factors in the development process. To guide the development of our study, we seek to answer the following research questions:

- RQ1: How do individuals perceive the interdependence of their tasks in software development?

- RQ2: How are perceptions of task interdependence related to other work-related factors and individual feelings in software development?

We believe answers to the above questions will be an important step towards a conceptual model of task interdependence in software development that reflects the system and process nature of interdependence in a teamwork context.

The rest of this paper is organized as follows. In Section II, we present the conceptual background and related literature. In Section III, we describe the research method used in this study. In Section IV, we present the results of our study. In Section V, we describe the implications of results for research and practice, and also discuss limitations and threats to validity. Finally, in Section VI, we present conclusions and recommendations for future work.

## II. BACKGROUND AND RELATED WORK

Work design can be defined as the set of opportunities and constraints in structured tasks and responsibilities that are allocated to the employee that affect how they perform and experience the work [12]. Morgeson and Humphrey [2] conducted an extensive integrative work to build a comprehensive model of work design, named the Work Design Questionnaire (WDQ). They identified 107 work characteristics terms in the literature, and grouped them in a classification model with 21 categories, organized as follows:

- Motivational characteristics are concerned with how the work itself is accomplished. It includes task characteristics, namely autonomy (split into work scheduling, decision-making and work methods), task variety, task significance, task identity and feedback from job, and the knowledge demands that are placed on an individual as a function of what is done on the job, including job complexity, information processing, problem solving, skill variety, and specialization;

- Social characteristics concern interpersonal and social aspects of work [14], such as social support, interdependence (initiated and received), interaction outside the organization, and feedback from others;

- Contextual characteristics reflect the physical and environmental context, namely, ergonomics, physical demands, work conditions, and equipment use.

The benefits of social work design characteristics, in particular in the context of teamwork and knowledge-based work, have been reaffirmed in an extensive meta-analysis [15]. Task interdependence is one such social job design characteristic, which is the focus of our study.

### A. Conceptualization of Task Interdependence

There are different conceptualizations of job interdependence that try to capture different types of interactions within and between levels of the organization, and also between organizations [16][29]. A thorough review of the literature about interdependence is outside the scope of this article and we refer the reader to the works of Wageman [16], Campion et al. [17], Van De Ven et al. [18], and Courtright et al. [19] for an overview of the literature.

Our goal is to study the interdependencies that happen between members of a software team related to software development tasks. Consistently, we use the conceptualization of task interdependence developed by Kiggundu [1] and operationalized by Kiggundu [20] and Morgesen and Humphrey [2]. In particular, Kiggundu [1] strengthened Hackman and Oldham's [12] job design theory by including task interdependence and hypothesizing that, together with autonomy, it is also an antecedent of "the critical psychological state of experienced responsibility for work outcomes" [1]. Considering the increase in the importance of individual autonomy, in particular in the context of agile software development [13], we contend that understanding the perception of task interdependence in a software team context has important implications for research and practice.

According to Kiggundu [1], task interdependence is related to the connectedness of workflow between tasks, in which the tasks are associated in such a way that the performance of one task depends on the effective performance of others. Kiggundu [1] also distinguished between initiated and received task interdependence. The former happens when the successful outcome of a task is required by one or more tasks that happen later in the workflow. The latter is defined as the extent to which a task is affected by the outcome of one or more tasks previously performed in the workflow.

Important to the practical implications of task interdependence, is the potentially opposing effects of initiated and received interdependence on individual motivation, satisfaction, and overall job outcomes. On the one hand, initiated task interdependence complements the effect of autonomy on increasing experienced responsibility, which leads to increased motivation [1][21]. On the other hand, received interdependence tends to lead to decreased autonomy, which in turn reduces motivation and satisfaction [1].

Other authors have found that there are interactive effects between task interdependence and autonomy, and that tradeoffs between these design factors occur naturally in creative teams [3]. Moreover, Morgesen and Humphrey [2], in a study with professionals from several different sectors, found that task interdependence correlated with several other work design factors, in particular with information processing (the degree to which a job requires attending to and processing data or other information). This suggests that the more interdependent the tasks are, the more they demand information processing for their development.

### B. Task Interdependence in Software Engineering

In this section, we present a brief review of task interdependence studies in software development. Acuña et al. [22] attempted to identify any type of relationship between software development team satisfaction on one hand, and autonomy and task characteristics, such as interdependence, on the other. They found that satisfaction is positively related to task interdependence as well as to autonomy (consistent with Kiggundu [1]), and that conflicts between team members and task conflicts negatively impact satisfaction.

More recently, Silva et al. [9] found evidence that interdependent tasks correlate:

- positively with the perception of role conflict in software teamwork. Together with the results from Acuña et al. [22], these findings suggest that properly managing interdependence in software development is important for individual job satisfaction;

- negatively with role interchangeability, suggesting that when software engineers are aware of an interdependence, they also perceive their role to be more difficult to be performed by others, which in turn can make it difficult to perform job rotation.

Kuthyola et al. [23] found strong evidence, while studying the impacts of interdependence in software development projects, that: task interdependence is positively related to teamwork quality; and the more interdependent the tasks, the stronger the relationship between teamwork quality and project performance.

Finally, Marsicano et al. [10] identified that in software development task interdependence increases the need for information sharing and synchronization of tasks. This is consistent with findings in other fields which showed that task interdependence correlates positively with information processing [2].

Some of these findings suggest that task interdependence is a work design characteristic that might have positive effects on motivation and satisfaction in software development, supporting the findings of Kiggundu [1] and Morgesen and Humphrey [2]. However, some findings also show potential negative effects related to an increase in managerial effort (synchronization of tasks and role conflict resolution) and the need for more effective information processing and communication management.

*C. Other Types of Interdependence*

Beyond task interdependence, other types of work interdependence exist in teamwork. Pennings [29] studied role (the position of a team member with respect to the others), social (the interdependent needs or goals of team members), and knowledge (the different skills and expertise of team members) interdependence in complement to task interdependence. More recently, Milhiser et al. [27] used these four components of work interdependence to study effective team building policies that take interdependence into account.

Although the other three components of work interdependence are also important in software development, in this initial study we focus on task interdependence. In future work, we plan to expand our studies to include role, social, and knowledge interdependence.

## III. METHOD

In this study, we are interested in understanding how individual software engineers interpret their experiences with task interdependence in the workplace and how these interpretations relate to psychological states that may positively or adversely affect work outcomes. Consistent with the nature of our research questions and investigated phenomenon, we performed an exploratory qualitative case study of a single software development team in an industrial context, following guidelines described by Merriam and Tisdell [24]. Quantitative measures of task interdependence are not in the scope of our study. The interested reader is referred to Kiggundu [20] and Morgesen and Humphrey [2] for operationalizations that measure the perceptions of individuals regarding the interdependence of their work with others.

*A. Selecting the Software Company*

We purposefully looked for a software company with the following characteristics:

- A mature company with over 15 years in the market.
- Software development was performed in teams and there were at least two development teams working simultaneously in different projects.
- Software teams were composed of members with a mix of roles dedicated full time to the team: requirement analyst, developers, testers, UX/UI designer, etc.

We did not impose restrictions on the type of technology used or the market targeted by the company. We also looked for companies in which we could have access to team members for data collection. Finally, we looked for a company in which management (of technical and/or human resources) would be interested in the investigation of the phenomenon of task interdependence.

The chosen case was a Brazilian software company that developed solutions for the financial market. During the study, the company had a total of 44 employees and three projects were running simultaneously. Project managers expressed their interest in the study because they had experienced conflicts in certain projects that they could attribute to the lack of mutual understanding of interdependencies between team members.

*B. Selecting the Software Team*

We discussed with project managers the best balance between the characteristics of the team and the level of workload the team was experiencing at the time of the research, because we sought to disturb their work as little as possible. We agreed to study a team with relevant characteristics for this study (Table I).

TABLE I.  TEAM CHARACTERISTICS

| Id | Gender | Time in Company | Time in the Team | Role |
|---|---|---|---|---|
| E01 | Male | 14 years | 4 years | Manager/Req. Engineer |
| E02 | Male | 4 months | 4 months | Developer - Front end |
| E03 | Male | 2 months | 2 months | Developer - Back end |
| E04 | Male | 8 years | 2 years | Test Analyst |
| E05 | Male | 2 months | 2 months | UX/UI designer |
| E06 | Male | 10 years | 6 years | Developer - Back end |
| E07 | Female | 2 months | 2 months | Developer - Full Stack |
| E08 | Male | 7 years | 3 years | Developer - Full Stack |
| E09 | Male | 20 days | 15 days | UX/ Web designer |
| E10 | Male | 3 years | 1 year | Developer - Full Stack |

The team was composed of 10 individuals in different roles; a good mix of company time (between 14 years and less than a month) and time in the project; and the team was not facing any crisis regarding deadlines or deliverables. The team consisted of 9 men and 1 woman. Although one could argue that a more balanced mix of gender would be preferable to increase diversity of perceptions, we could not study other teams due to the company's restrictions.

*C. Data Collection*

Semi-structured interviews were conducted with all 10 members of the software team. The interview script was built following the recommendations of Merriam and Tisdell [24]. We validated the interview scripts by conducting pilot interviews with a group of five professionals from different companies, with 2-6 years of work experience. We made minor adjustments to the phrasing of some questions and also timed the pilot interviews to have an estimate of the duration of the actual interviews. We estimated 20-30 minutes for the interviews. Fig. 1 shows examples of the questions asked in the interviews and the full interview script is presented in Appendix A.

| 18. How do you feel when a team member depends on you to carry out her/his activity? |
| 20. What information do you need to carry out your tasks? |

Fig. 1. Examples of Interview Questions

The original interview script was written in Portuguese and the interviews were conducted in the same language. We translated the interview script to present it in this paper.

All interviews occurred in the organization's facilities and were performed by an interviewer and supported by a second researcher (who took notes to support data analysis). All interviews were recorded and lasted an average of 20 minutes each. Altogether, they produced nearly four hours of audio and over 50 pages of transcriptions.

*D. Data Analysis and Synthesis*

Initially, all audio from the interviews was verbatim transcribed. The transcriptions were produced in Portuguese and the data analysis was performed in Portuguese as well. Translation of quotes, codes, and concepts was carried out after the analysis was completed to be presented in this article.

We used standard coding techniques from qualitative research to code, categorize, and synthesize data, with the support of MAXQDA Analytics Pro version 12. Data analysis was performed in two iterations. The first looked for answers to research questions RQ1 and RQ2. The results of this iteration are presented in Section IV-B. The second iteration was prompted by findings of the previous iteration and the results are presented in Sections IV-C,D.

Each iteration began with open coding of the transcripts. Postformed codes were constructed as the coding progressed and were attached to particular pieces of the text. Then, the codes arising from each interview were constantly compared to codes in the same interview and from other interviews, resulting in categories that represented concepts related to task interdependence (Fig. 2).

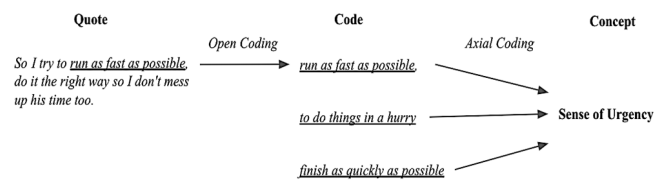

Fig. 2. Open and Axial Coding: Building Concepts

As the process of data analysis progressed, relationships among categories were found. In Fig. 3, we illustrate how we constructed a relationship between sense of urgency and anxiety.

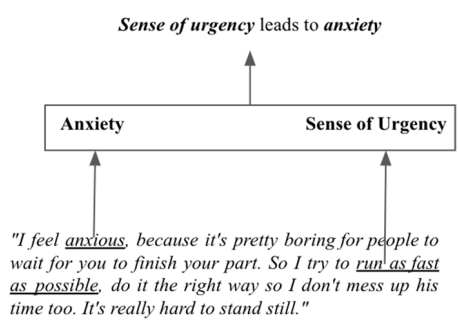

Fig. 3. Building relationships

*E. Ethics*

We followed the norms of Resolution 466/12 – CNS-MS [26] of the Brazilian National Health Council, which regulates research with human subjects. The company signed a Term of Authorization and the researchers signed a Non-disclosure Agreement (covering access to sensitive information). Both documents granted the researchers access to facilities, to the participants, and to necessary documentation. They also authorized the participants to use work hours for the interviews. We believe that this formalization reduced the possibility of participants concealing information that they would consider sensitive. Before the interviews, the participants agreed to the terms of an Informed Consent Form that explained the overall objective and relevance of the research, guaranteed data confidentiality, the anonymity of the participation, the non-obligatory nature of the participation, and the right to withdraw from the research at any moment. All invited individuals freely agreed to participate and no participant withdrew from the study.

## IV. RESULTS

We start with a description of the company and participants. Next, in Section IV-B, we describe the dynamics of task interdependence in the work of the software team, obtained from the perceptions of the participants. These findings led us to investigate symmetry in task interdependence relationships. In Section IV-C, we explore this issue, and in Section IV-D, we show how asymmetric perceptions of task interdependence may impact individual emotional states.

*A. Context Description*

In this section, we briefly describe the context of the company and the software team we studied.

The software company was founded in 1989 and, at the time of the study, had 44 employees. The company developed software for the financial market, focusing on products related to factoring, credit and collection, securitization, Credit Rights Investment Funds (FIDC), and digital signature, among others.

The software development was organized in three teams:

- Team 01: 1 manager and 6 developers, focused on the maintenance and evolution of products that use Windows™ systems.

- Team 02: 1 manager and 4 developers, focused on the maintenance and evolution of web products.

- Team 03: 1 manager and 9 developers, focused on innovations and maintenance.

The chosen team (Team 03) used Scrum to organize and manage the development process. Team 03 focused on the development of new products, and on evolutionary and corrective maintenance of products in the financial area. Daily meetings were held for 15 minutes, in which the team assessed the progress of its ongoing project. Tests were performed at the end of each sprint to confirm that the developed product met the established requirements. At the time of the research, the team was working remotely (home office) due to the COVID19 pandemic, using the following tools: Meeting, Google Docs, Trello, NodeJS, Figma, PostgreSQL, Angular, Monday, and the Java and Javascript programming languages.

### B. Task Interdependence and Correlated Factors

Searching for answers to the research questions RQ1 and RQ2, we analyzed the interview data looking for the perceptions of the team members regarding task interdependence and other factors, such as other work characteristics and personal feelings, that were correlated with interdependence.

*1) Characterization of Task Interdependence.* Initially, we coded the interviews to synthesize a characterization of task interdependence as perceived by the team members. We found clear recognition of task interdependence, in which team members were aware that tasks performed by different individuals depended on each other.

*"If I don't finish the demands that are about to be tested, they won't be able to upload the next group of demands and that slows down the process a bit." [E04]*

*"Because everything I do I need a dev to code, so necessarily if I don't do my job, the dev doesn't do his. So there's an interdependence of me producing for the dev to produce and the dev has to produce for me to do what I'm doing on the screen, so this is full interdependence." [E09]*

Consistent with our conceptual background, depending on the direction of the workflow in relation to task owner, we found perceptions of Received Interdependence (the task that is unable to begin until another task is complete):

*"Sometimes, one part depends on the conclusion of another one for the two to bind together." [E07]*

*"Sometimes I get idle because I depend on the design and requirements to get things done." [E03]*

We also found perceptions of **Initiated Interdependence** (the task whose completion allows the next one to begin):

*"I'm doing it to help the next activity, collaborate with the next activity." [E01]*

In summary, team members identified interdependent relations in their work; perceptions of initiated and received task interdependence were found; team members were aware that, in interdependent tasks, the performance of one task depended on the effective performance of others and vice-versa. We therefore concluded that the perception of task interdependence in this software team was consistent with the conceptual background presented in Section II.

*2) Work Design Characteristics.* As discussed in Section II-A, studies in other fields identified that task interdependence correlates with other work design characteristics. We then continued our analysis to search for other work design characteristics experienced by the software team that could interact with task interdependence.

We interpreted that the team members experienced low levels of autonomy in all three facets defined in the WDQ [2]: work scheduling, decision making, and work methods autonomy.

*"We have a panel where we have the activities to be done and we see the priorities of these tasks and we do it." [E10]*

*"Today we have a tool, MONDAY, which has all the steps, very well detailed, and those responsible for each activity. So, I have a queue that is mine, which whenever someone finishes and goes up to my approval environment, it automatically appears in my queue. I run tests." [E04]*

We also found that, at the individual level, the way teamwork was designed produced low levels of task variety (the extent to which the work required team members to perform a wide range of tasks [2]). This was clear in the way roles were defined:

*"I am a test analyst. ... There are developers, each one in a different aspect like the front end, back end." [E04]*

*"I work as an interface and user experience designer." [E05]*

The literature shows that autonomy and task variety are positively correlated [2]. Thus, our findings of low autonomy and low task variety are consistent with the theories of work design. The low levels of autonomy, in particular related to the work scheduling facet, and task variety led us to contend that:

**Proposition 1:** Work design characteristics of **autonomy** and **task variety** moderate the effect of task interdependence on **managerial effort** (synchronization of tasks and role conflict resolution).

We found that most team members perceived high levels of social support (the degree to which a job provides opportunities for advice and assistance from others [2]) in their job.

*"Our team is very tight-knit! ... if I have a problem, I ask them, they help me and vice versa." [E02]*

*"So, sometimes, for example, I'm also having a hard time finishing something, I ask a colleague for help and he helps me." [E07]*

*"But in my experience, I don't take a task and I'm going to do it and carry it out until the end. Usually, I talk to someone. I go and talk to the requirements analyst to ask how this thing is going to work, ... we usually talk to other people to understand what should be done." [E10]*

This finding led to the following proposition, consistent with the findings from da Silva et al. [9], who found positive correlations between social support and task interdependence:

**Proposition 2: Social support** moderates the effect of task interdependence on the need for **information sharing** within the team.

*3) Social Characteristics of the Work.* We continued our analysis looking for social characteristics of teamwork that could be related to task interdependence. We found that the awareness of interdependence between tasks was related to the perception that team members should develop and strengthen effective **interpersonal relationships**. The more they perceived their tasks as being interdependent, the more the completion of the tasks would benefit from these relationships.

*"... we try not to put any obstacles in the way. We are always negotiating things." [E01]*

*"When I depend on another developer, I call him to a meeting. I tell him what we need to do. We agree on the responsibilities and divide the problem ... and everyone does their part and then consolidate them together." [E02]*

*"And as a practice, I finish everything by talking to the person in charge directly. ... I say: "look, I finished all the demands. You can now go up to production and release the next ones". In addition to the system, I still do the communication so that I can speed up the process and we can move on to the next activity." [E04]*

Also, team members involved in interdependent relationships put an extra effort to achieve collective task synchronization as much as possible.

*"... I think we make the greatest effort to walk together in these tasks that are dependent on each other." [E10]*

These finding led to the following proposition:

**Proposition 3:** Effective **interpersonal relationships** will facilitate team communication and **collective task synchronization** of interdependent tasks.

*4) Feelings Related to Task Interdependence.* We then analyzed what feelings were expressed by the team members with respect to working on interdependent tasks. Consistent with the literature, initiated and received interdependence produced different feelings.

    *a) Initiated Interdependence*

Some team member expressed that their awareness of being at the initiating end of an interdependence created feelings of increased **responsibility** for delivering their tasks in timely way and with good quality, and that to accomplish these deliveries they also improved their **self-organization**:

*"It's a matter of responsibility, I feel responsible." [E05]*

*"... over time I understood, even knowing that others depend on me, if I take it easy and organize myself better ... I don't see a problem." [E03]*

One team member expressed that starting the workflow in an interdependent relation was **challenging** and perceived that challenge as a positive feeling.

*"The feeling is the best possible, I don't have the pressure on me. I think it's very challenging, but I don't see it as pressure, it's not something that bothers me. I find it wonderful and challenging." [E09]*

The feeling most often expressed by those that were on the initiating end of the interdependence relationship was a **sense of urgency** to finish their tasks so as not to disrupt the workflow. This was expressed by nearly half of the team members.

*"I get a little uncomfortable, wanting to do things in a hurry. Because like it or not, he gets a little idle." [E02]*

*"Finish my tasks as quickly as possible so I don't get in the way of other people's tasks." [E04]*

For some team members, this sense of urgency was not perceived as negative, instead a potential motivator for increased commitment to the task, as for E02 above. For others, this feeling was associated with anxiety:

*"I feel anxious, because it's pretty boring for people to wait for you to finish your part. So, I try to run as fast as possible, do it the right way so I don't mess up his time too. It's really hard to stand still." [E07]*

Finally, some team members expressed frustration when they perceived that their initiating end of the task interdependence did not produce timely or quality results for the receiving end.

*"... you're going to feel a little frustrated, when a person depends [on me] and I don't pass the activity with documentation of good perception that he understands ..." [E01]*

Overall, team members on the initiating side of the interdependence expressed a mix of positive and negative feelings, summarized in Table II.

TABLE II.     FEELINGS RELATED TO INITIATED INTERDEPENDENCE

| | |
|---|---|
| + | **Responsibility leading to improved self-organization.** |
| + | **Challenging but not distressing.** |
| + | **Sense of urgency increasing the commitment to the initiating task.** |
| - | **Sense of urgency leading to anxiety.** |
| - | **Frustration due to low quality of initiating task results.** |

    *b) Received Interdependence*

When on the receiving side of an interdependent relationship, team members expressed mostly negative feelings. For some, received interdependence impacted their workload by either increasing (work overload) or decreasing (work underload) the work to a substantial degree that impacted their productivity.

*"... it creates emergencies! Like: "I have this to do and I still haven't received it!". Sometimes I get idle **[work underload]** because I depend on the design and requirements to perform the tasks. But right now, I'm trying to learn more about the requirements and I feel like I have a lot of work **[work overload]**." [E03]*

*"If I don't have the requirement, I don't know what I need to do **[work underload]**." [E05]*

As a consequence of this impact on workload (in particular, related to a sense of work underload), team members expressed a feeling of distress related to the incapacity of doing their work due to the lack of results from the initiating end of the interdependence.

*"So, I feel like this: distressed. Because I know there's a lot to be released, but I can't do anything until it's released for me to do. Oh, I feel a little nervous. Like: "people, free up [the tasks] for me to work." [E04]*

Other team members on the receiving end of an interdependence reported feelings of **anxiety** when waiting for inputs from the initiating tasks.

*"I depend on information from developers and programmers to be able to produce my work. I get anxious waiting for the next demand." [E04]*

Also associated with the sense of work underload was a feeling of frustration. In this case, different from the **frustration** due to the low quality of the initiated task (Initiated Interdependence), here the frustration was for not having inputs to perform the receiving end of the interdependent task.

*"I feel a little frustrated because it turns out that I can't do anything until he's finished. I end up going to study something or I'm going to help them on the case." [E08]*

In Table III, we summarize the feelings related to the perception of being at the receiving end of an interdependent relation.

TABLE III. FEELINGS RELATED TO RECEIVED INTERDEPENDENCE

| - | Work overload due to extra work to compensate for lack of input on the receiving task. |
|---|---|
| - | Work underload due to lack of inputs on the receiving task. |
| - | Anxiety when waiting for inputs from the initiating tasks. |
| - | Distress due to negative impact on workload. |
| - | Frustration related work underload due to lack of inputs on the receiving task. |

### C. Symmetry of the Perceptions of Task Interdependence

While analyzing the data looking for perceptions related to task interdependence, we identified that certain team members did not express any awareness of being in the initiating end of an interdependence. For instance, when E02 (front-end developer) was asked if there were other tasks in the team that depended on his work, he answered:

*"I do not think so. Let me think... No, not that I know of. I don't think so. That question got me thinking."[E02]*

This raised the suspicion that perceptions of interdependence relationships were not symmetrical between the two ends of the relationship. Considering that the feelings related to the two types of interdependence (Tables II and III) are very different, with initiated interdependence being related to more positive feelings that received interdependence, we contend that asymmetrical perceptions of interdependence will exacerbate negative feelings and attenuate the positive ones. So, investigating this asymmetry seems important.

We then reanalyzed the interviews looking for the symmetrical and asymmetrical perceptions of each pair of team members involved in interdependent relationships. We found three types of perceptions:

- symmetrical: initiating and receiving task owners both had a perception that their tasks were interdependent.
- asymmetrical: one of the owners had a perception of interdependence from both sides and the other only viewed the interdependence from her own perspective.
- highly asymmetrical: one of the owners had a perception of interdependence from both sides and the other did not express any perception of the interdependence in that relationship.

*1) Symmetrical Perceptions.* In some relationships, we found that both sides shared the awareness of the interdependence between their tasks, agreeing that initiating and receiving tasks depended on each other for their effective development. For instance, team leader E01 produced the business requirements to designer E05 who produced the design, which then had to be approved by E01 for the task to be completed. Both sides knew that E01 can only continue sending requirements to E05 after approving the previous one.

*"I bring the demands... I myself plan what will be studied. Then ... the designer E05 takes his activity and produces the design. After my approval of the design, the developer already takes..." [E01]*

*"E01 ... brings the requirements when we are going to develop some functionality... I make some screen sketches, some proposals to deliver a solution. From that moment on, I move on to validation with E01." [E05]*

Similarly, developer E08 shared the same perception with E04 that the workflow between development and test were interdependent.

*"If I don't finish the demands that are about to be tested, they won't be able to upload the next groups of demands and that slows down the process a bit" [E04].*

*"I generate an approval to release for the tester, after the tester releases, I generate a production environment." [E08]*

*2) Asymmetrical Perceptions.* We also found instances of asymmetrical perceptions of task interdependence, in which one side perceived the relationship as an interdependence and the other side perceived as a dependence on a single direction. For instance, leader E01 was aware of the interdependence he had with designer E09, similarly as how he expressed about the relationship with E05, described above. However, although E09 expressed his perception as being on the receiving end of the relationship:

*"I wouldn't be able to start if there wasn't a demand from the customer. So, this need comes from E01 until it gets to me." [E09]*

When asked about his perceptions of being on the initiating end of the relationship, E09 did not express that his activity needed the approval from E01 for the whole task to be completed. In fact, he expressed that:

*"Today, I want to make it clear that there is still no such interdependence, but these are things that we have been aligning." [E09]*

*3) Highly Asymmetrical Perceptions.* The second type of asymmetry of perceptions was called highly asymmetrical because while one side was fully aware of the interdependence relationship, the other did not express any perception of the relationship.

For instance, the front-end developer E02 and the full-stack developer E10 did not express any perception as being part of an interdependent relationship with tester E04. Only E04 perceived the initiated and received interdependence. In fact, they did not perceive being part of any initiated interdependence, as expressed by E02 in the quote that prompted our investigation of the asymmetries, already presented above. Similarly, developers in general did not express having a perception of task interdependence with team leader E01.

In summary, we found three situations regarding the symmetry or asymmetry of perceptions of interdependence in this team. Mostly, the team leader, the designers, and the test analyst were aware of both sides of the interdependence, at least in some of their relationships. On the other hand, developers were the source of most of the asymmetrical perceptions. We shall discuss these findings in the next section.

*D. Causes and Consequences of Asymmetric Perceptions*

We then reflected on potential causes of the asymmetric perceptions by reanalyzing the findings related to work design characteristics, which were discussed in Section IV-B. These findings indicated that the team experienced low levels of autonomy due to the rigid way in which tasks were defined (by the team leader) and allocated to other team members and, in particular, the developers. Also, team members experienced high levels of work specialization, in particular developers, who were clearly specialized in front-end and back-end. These work design characteristics led us to the following proposition regarding possible causes of the asymmetric perceptions of task interdependence.

**Proposition 4**: Low levels of **autonomy**, implemented by a rigid command and control management, will hinder the visibility of task interdependencies, potentially creating asymmetries of perceptions between the task owners involved in the interdependent relation, which will be exacerbated by high levels of job **specialization**.

We then looked for the potential consequences of these asymmetries. One team member involved in all three types of situations was the test analyst E04. He was also the one that most emphatically expressed that the work underload on the receiving end of the interdependence was related to his feelings of distress and anxiety. Which led us to the following proposition:

**Proposition 5:** The feelings of **anxiety** and **distress** due the **work underload** on the receiving side of the interdependence relationship is exacerbated in the presence of asymmetric perceptions of the interdependence.

In summary, we contend that asymmetric perceptions of interdependence are related to work design characteristics of the tasks and, in particular, low levels of autonomy and high levels of specialization. Also, we contend that the negative feelings experienced by those on the receiving side of the interdependence relationship are exacerbated by the asymmetric perceptions of those in the initiating side of the relationship.

## V. DISCUSSION

We studied the interplay of task interdependence and other work design characteristics from the perspective of software engineers in a teamwork environment. We also studied how team members expressed their feelings about working on the initiating and receiving ends of an interdependence relationship.

In work design theories, task interdependence is a motivating factor because, together with autonomy, it is an antecedent of "the critical psychological state of experienced responsibility for work outcomes" [1]. Therefore, work designs with high levels of interdependence and autonomy should lead to higher commitment and, thus, motivation [21].

However, designing software development work with high levels of both interdependence and autonomy is not straightforward. In fact, Kakar [4] found that high levels of task interdependence are likely to decrease individual autonomy. Furthermore, previous research has shown that task interdependence in software development is associated with increase in **managerial effort** due to an increased need for synchronization of tasks [10] and role conflict resolution [9]. Also, highly interdependent work requires more information processing [2] and information sharing [10]. Moreover, high levels of interdependency between software team members may be perceived as counterproductive because the time spent on negotiations and decision making processes could be used to carry out the actual development tasks [4][11].

Our research has added to these findings, showing that certain work design characteristics may reduce the potential negative effects of task interdependence:

- Low levels of autonomy and task variety may reduce the need for extra managerial effort in interdependent tasks because this type of work design reduces negotiations and decision making at the team level (Proposition 1).

- Levels of social support between team members may be increased by the awareness of task interdependence (Proposition 2).

- High levels of social support may reduce the need for extra managerial effort in interdependent tasks due an increase in direct information sharing (Proposition 2).

- Effective interpersonal relationships will facilitate team communication and collective task synchronization of interdependent tasks (Proposition 3).

Nevertheless, our findings also suggest that low levels of autonomy (in particular, working schedule autonomy) and high levels of specialization may create asymmetric perceptions of task interdependence (Proposition 4), which, in turn, may exacerbate feelings of anxiety and distress for team members on the receiving end of (asymmetrically perceived) interdependence relationship (Proposition 5).

*A. Implications for Practice*

Choices of work design characteristics, such as autonomy, task interdependence, and task and skill variety, may have significant impact on motivation and satisfaction in practice. Our study showed that these characteristics may complement or compete with each other. For instance, while high levels of autonomy and interdependence may create a highly motivating working environment, they also will increase the need for more information sharing and social support, and potentially create role conflicts.

Although our results are based on the study of a single software team, it is consistent with the literature on work design and also with previous research on software teams discussed in Section II. Therefore, we believe that managers and team leaders could use the propositions and, in general, the findings of this study as a guiding framework for managing task interdependence in practice. Considering that technical system interdependence may not be possible to completely avoid, focus on the management of social, process, and organizational interdependencies by considering that:

- Low levels of autonomy and task variety reduces the impact of task interdependence, but may create asymmetric perceptions of the interdependence relationships between team members.

- Teams with high levels of social support between team members and effective interpersonal relationships will be able to reduce the effects of asymmetric perceptions through the increase in information sharing. These social characteristics of the team should be fostered.

- Asymmetric perceptions of interdependence should be managed effectively to reduce the feelings of anxiety and stress mainly in the receiving end of the interdependence relationship.

We believe that the main takeaway of our study for industry is that work design choices, which occur naturally in the workplace, should be made by carefully analyzing the potential interplay of the various work design characteristics, as discussed above. Our study presented some of these interactions in a way that we hope can raise the awareness of managers and team members about their potential benefits and drawbacks. In particular, our study shows that asymmetries in the perception of task interdependence may exist in practice and are likely to increase negative feelings in the workplace. We believe these asymmetries should be identified and mitigated.

*B. Implications for Research*

As we discussed in the Introduction, software development entails high levels of task interdependence, created by the technical dependencies between the components of the software system (system interdependence) and also by how the development tasks are allocated to individuals in a teamwork context (process interdependence). In this study, we explored some aspects of task interdependence related to the way tasks are allocated and developed, which is part of what we call process related interdependence. We believe this study has to be extended to address other work design characteristics. Furthermore, we need to better understand the interplay between task interdependence from a process perspective with system interdependence. In this direction, one potential research topic would be to investigate the relationship between interdependence and technical debt, as the decisions regarding acquiring, paying, or deferring a debt may be related to levels of task interdependence.

Also, besides task independence, research in organization science has addressed other components of interdependence: "role (the position of the respective team members relative to each other), social elements (the mutual needs or goals of members), and knowledge (the differentiated expertise of the members)" [27][29]. We believe that these three components are relevant in the composition of software engineering teams, as well as in the decisions regarding work design characteristics in software development.

*C. Addressing Limitations, Validity and Reliability*

Validity and reliability assessments used in positivist experimental studies do not apply directly to interpretive qualitative research. We discuss the validity and reliability of our results from the perspectives proposed by Merriam [28].

Construct validity in qualitative research is related to the precise and clear-cut definition of constructs that is consistent with the meanings assigned by the research participants. Our analysis was conducted by two researchers, to minimize potential misinterpretations. We also compared and contrasted our construct definitions with the literature. A further step we could have taken to ensure construct validity was to formalize our definition of task interdependence, define a consistent protocol for providing participants with this definition, and test their understanding of it.

Internal validity, or credibility, is related to the extent that the results match reality and that the researchers were able to capture reality as closely as possible. To increase credibility, we tried to achieve maximum variation on the sources of data (team members) within a single software team. We collected direct data from all team members, so no perspective would be left out. We then compared the findings with the literature on software engineering and other fields to sharpen construct definitions and increase internal validity.

Reliability refers to the extent to which the results can be replicated. We do not expect results from qualitative research to be replicable in a positivist sense because, in short, human behavior, feelings, and perceptions change. The more important question is whether the results follow consistently from the data and that the researchers did not make any inference that cannot be supported by the data, which Merriam refers to as consistency [28]. To increase consistency, two researchers performed the data analysis independently. Then, the second author merged the results. A few inconsistencies were discussed in a meeting. The third author reviewed the findings. Future studies could incorporate member checking to further ensure reliability.

Our results provide a rich description of the studied phenomenon, which may be transferable (instead of generalizable). In this sense, although we do not expect that all our findings will be applicable to other contexts, it is possible to learn from the case description and decide to what extent the findings can be adapted and/or transferred to other situations (we hope that the discussion in Section V-A will assist managers to do this transfer). Two strategies were employed to enhance transferability of the results. First, we tried to provide a rich description of the research method, the context in which the research was performed, and the results themselves. However, we believe this is one of the limitations of this article, since space limitations prevent full reporting of rich and thick descriptions of the context. Second, we interviewed all team members to achieve maximum variation possible within the team because this would provide richer data and, consequently, richer results more widely applicable.

This study investigated a single software team of 10 team members in a single software company. One common (although misguided) criticism about single case studies is that they are too small to make conclusions. Flyvbjerg [30] considers this criticism one of the five big misunderstandings about case study research. In his view (which we also share), conceptual generalization (and not statistical generalization) is not only possible from single case studies but is an essential part of knowledge development in all sciences. Further, this criticism is also in the list of "invalid criticism" of the ACM Empirical Standard [31]. There is a lot to be learnt from single case studies, as we hope to have shown with the discussions above.

## VI. Concluding Remarks

In this article, we presented the results of a case study about task interdependence between software team members working in a software development company. Initially, we investigated the perceptions of team members about interdependence of their tasks, how interdependence related to other work design characteristics of their job, and how interdependence was perceived as affecting emotional states and feelings in the workplace. Prompted by findings from the initial analysis, we also investigated the symmetrical and asymmetrical perceptions of task interdependence between the initiating and receiving end of the interdependence relationships.

We found that team members are consistent in their conceptual understanding of what task interdependence is and how it happens in practice. Also, consistent with the literature from other fields, we identified that task interdependence interacts in complex ways with other work design characteristics, in particular with autonomy, and that these interactions might be moderated by levels of task variety and job specialization.

Moreover, tasks interdependence is associated with emotional states and feelings in the workplace that might impact motivation and satisfaction at work. Consistent with the work of Kiggundu [1], initiated and received task interdependence are associated with different feelings, with team members on the initiated end of the interdependence relationship expressing more positive feelings than those on the receiving end.

Finally, we identified that perceptions of task interdependence are not symmetrical between all team members. Some team members, mostly on the initiating end of the interdependence relationship, failed to express their perception of the relationship with the receiving end. We contend that such asymmetries may exacerbate feelings of anxiety and distress expressed by those in the receiving end of the relationship. Due to its potential negative effects on teamwork effectiveness, these asymmetries should be explicitly managed in practice.

We then propose future research in two directions. The first direction is to study task interdependence in the context of system interdependencies that are created by how the software system is architected and how the parts are divided and allocated to the software teams. The second is to expand the study of interdependence in software development to other components such as role, social, and knowledge interdependence, as discussed by Pennings [29] and also addressed by Milhiser et al. [27].

One of the limitations of this study is that it was conducted in a single company and in a single software team. Thus, we do not expect the findings to cover all possible interpretations of the phenomenon nor to be immediately generalized to other contexts. Nevertheless, the consistency of our findings with the extant literature increased our confidence on the validity and consistency of our results, and also its potential transferability (with possible adaptations) to other contexts. We hope this study stimulates the empirical software engineering research community to pursue other studies in this topic.

## Acknowledgments

The authors would like to thank the participants of this study. Prof. Fabio Q. B. da Silva receives a research grant from CNPq 303738/2020-0.

## Appendix A - Interview Script

**Introduction**

1. Introduce the interviewers; explain the purpose of the interview; ask permission for recording audio.

**Background and Context**

2. What is your academic background?

3. How much experience do you have in software development?

4. How long have you been working at this company?

5. How long have you been on this project?

6. What is your current role on the team?

7. Describe the roles of the other members of your team.

8. Tell me a little about your typical work day.

9. Do you feel part of a team?

10. How is the division of tasks in your team?

**Task Interdependence**

11. Is it common for your team's tasks to depend on the involvement of more than one person to be completed?

12. How does this task allocation work?

13. In your opinion, does this allocation work properly?

**Initiated Task Interdependence**

14. Are there tasks that can't be carried out before you finish yours? How do you feel about it?

15. What information do you generate so that other people can do their work? How do you generate this information?

16. How do the results generated by your tasks affect the performance of the rest of the team?

17. How do you feel when a team member depends on you to carry out her activity?

**Received Task Interdependence**

18. Are there tasks you can't carry aout before others have finish their task(s)? How do you feel about it?

19. What information do you need from others to carry out your tasks?

20. Do the results generated by other members of your team interfere with your performance?

21. Could you say which people interfere (relate, generate information, etc.) directly in your work?

**Closing and thanks**

22. Any questions you'd like to add that haven't been asked here?

23. Thanks for participating in the survey.